\documentclass[aps]{revtex4}
\usepackage{eurosym}
\usepackage{amsfonts}
\usepackage{amsmath}
\usepackage{amssymb,epsf}
\usepackage{color}
\usepackage{graphicx}
\usepackage{epstopdf}
\usepackage{float}
\usepackage[font={footnotesize}]{caption}
\usepackage{subfig}
\usepackage{titlesec}

\begin{document}

\title{Effect of rainbow function on the structural properties of dark
energy star}
\author{A. Bagheri Tudeshki$^{1}$\footnote{%
email address: a.bagheri@hafez.shirazu.ac.ir}, G. H. Bordbar$^{1}$\footnote{%
email address: ghbordbar@shirazu.ac.ir}, and B. Eslam Panah$^{2,3,4}$\footnote{%
email address: eslampanah@umz.ac.ir}}
\affiliation{$^{1}$ Department of Physics and Biruni Observatory, Shiraz University,
Shiraz 71454, Iran \\
$^{2}$ Department of Theoretical Physics, Faculty of Science, University of
Mazandaran, P. O. Box 47415-416, Babolsar, Iran\\
$^{3}$ ICRANet-Mazandaran, University of Mazandaran, P. O. Box 47415-416,
Babolsar, Iran\\
$^{4}$ ICRANet, Piazza della Repubblica 10, I-65122 Pescara, Italy}

\begin{abstract}
Confirming the existence of compact objects with a mass greater than $%
2.5M_{\odot}$ by observational results such as GW190814 makes that is
possible to provide theories to justify these observational results using
modified gravity. This motivates us to use gravity's rainbow, which is the
appropriate case for dense objects, to investigate the dark energy star
structure as a suggested alternative case to the mass gap between neutron
stars and black holes in the perspective of quantum gravity. Hence, in the
present work, we derive the modified hydrostatic equilibrium equation for an
anisotropic fluid, represented by the extended Chaplygin equation of state
in gravity's rainbow. Then, for two isotropic and anisotropic cases, using
the numerical solution, we obtain energy-dependent maximum mass and its
corresponding radius, and the other properties of the dark energy star
including the pressure, energy density, stability, etc. In the following,
using the observational data, we compare the obtained results in two
frameworks of general relativity and gravity's rainbow.
\end{abstract}

\maketitle

\section{Introduction}

Although until now, by performing various observations, it is possible to
state the mass-radius relation and other properties of compact objects, the
main problem is determining the equation of state (EOS) governing the inner
fluid of the star, because we are not able to correctly determine its
nature. The recent accelerating expansion of the universe suggests the
existence of a cosmic fluid called dark energy, which is anti-gravitational 
\cite{Peebles2003}. In the structure of compact objects, the EOS of dark
energy can also be used as a definition of the inner fluid. The advantage of
this choice is that the singularity in the center is avoided. Various
stellar models such as false vacuum bubbles \cite{Coleman1980}, non-singular
black hole \cite{Dymnikova1992}, and gravastar \cite{Mazur2004} were
presented, each of them used different types of dark energy models in the
star structure. But the concept of a dark energy star (DES) was first
presented by Chapline \cite{Chapline} assuming the formation of dark energy
under a quantum phase transition within the star, visualized a compact
object with a quantum critical surface and a singularity-free center. In the
following, the stability of this compact object was studied in the presence
of anisotropy with the equation of state $p=\omega \rho $ \cite{lobo2006}.
This EOS with negative dark energy parameter $\omega $ creates pressure
against the gravity to prevent the gravitational collapse. By using this
EOS, several studies \cite%
{Yazadjiev2011,Bhar2015,Bhar2018,Beltracchi2019,Banerjee2020,Sakti2021}
about dark energy stars and their properties such as pressure behavior,
energy density, gravitational profile, and anisotropy were conducted in the
last two decades. However, the interest has moved towards the mixed EOSs,
which describe the inner behavior of a DES in the presence of dark energy
plus various matters. By combining dark energy and ideal Fermi gas, Ghezzi 
\cite{Ghezzi} presented a model for the EOS of a dark energy star, which was
able to estimate the maximum mass to about $2M_{\odot }$. An EOS including
dark energy and baryonic matter was used in Refs. \cite{Rahaman2012,Bhar2021}%
, where was shown in Ref. \cite{Bhar2021} that the maximum mass can reach
close to $2.5M_{\odot }$. A suitable option to describe the behavior of dark
energy is the Chaplygin gas (CG) with EOS $p=-\hat{B}/\rho $, which can
mimic the behavior of dark matter at early times and dark energy at late
times \cite{Kamenshchik2001,Zheng2022}. It was shown that by adding a
barotropic term to Chaplygin gas EOS, the maximum mass of the dark energy
star reaches more than $3M_{\odot }$ \cite%
{Panotopoulos2020,Panotopoulos2021,Pretel2023,Tudeshki2023}. These results
can open a new window to the mystery of the mass gap region $(2M_{\odot
}-5M_{\odot })$ \cite{Abbott2020,Ozel} between neutron stars (NS) and black
holes (BH). This means that dark energy stars can be a candidate for compact
objects that fill this mass gap region.

Today, several observational reports of the existence of binary star systems
have been obtained, which indicate the existence of a companion whose mass
can be placed in this mass gap region. Here, some candidates are GW170817 
\cite{Abbott2017} with total mass of approx $2.73M_{\odot}$, GW190425 with
the mass $3.4M_{\odot}$ \cite{Abbott2020L3}, and GW190814 with the mass $%
2.6M_{\odot}$ \cite{Abbott2020L44}. Also, the unseen companion of the binary
system with the giant 2MASS J05215658+4359220 with the mass $3.3M_{\odot}$ 
\cite{Thompson2019}, massive object in PSR J2215+5135 with the mass $%
2.27M_{\odot}$ \cite{Linares2018} and MSP J0740+6620 with the mass $%
2.14M_{\odot}$ \cite{Cromartie2019}, are also other possible options.
Although it is possible to formulate the structure of stars using
observational data in the framework of general relativity (GR), however, due
to its problems (for example, explaining the current acceleration of our
universe and in high energy regime), modified theories of gravity can also
be used for this formulation and, of course, to justify the observational
results.

In high-energy physics, and especially where the energy of particles can be
considered up to the Planck energy $E_{p}$, we need alternative gravity to
determine the effect of high energies on the formulation of the system. DES
introduced by Chapline \cite{Chapline} has a surface that allows quantum
phase transitions for spacetime. The particles are present at the surface
with high energy, which is transformed into dark energy under this phase
transition. The gravity's rainbow (RG) is a theory that is based on quantum
gravity and in the UV frequency (high energy) limit, it causes corrections
in the energy-momentum dispersion relation \cite{Magueijo 2004}. The
consequence of this modification is the dependence of the spacetime on
energy, which recreates GR in the low frequency (IR) limit. Energy-dependent
rainbow functions in RG can influence some properties of compact objects. In
this regard, the thermodynamic properties of BH in RG, such as entropy,
temperature, Hawking radiation, and other properties have been investigated
in several different studies \cite%
{Galan2006,Ling2007,Liu2008,Leiva2009,Ali2014,Ali2015,Feng2017,EslamPanah2018,EslamPanah2023}%
. Also, the thermodynamic properties of charged BHs in dilatonic gravity's
rainbow and their dynamic instability in RG have been studied in Refs. \cite%
{HendiFaizal2016,HendiPanahiyan2016}. It was shown that the neutron star
structure is dependent on rainbow functions \cite%
{HendiBordbar2016,EslamPanahetal2017}. Indeed, the maximum mass and radius
of neutron stars depended on rainbow functions. For the DES, it was found
that the rainbow functions increase the stable regions \cite{Tudeshki2022}.
With these interpretations, in this work, we intend to find the maximum mass
of the DES by studying the energy-dependent spacetime effects and
investigating its other properties under these conditions.

\section{Field and Modified TOV Equations in RG}

\label{sec2} In the framework of RG, the modified energy momentum dispersion
relation is defined as follows \cite{Magueijo 2004}%
\begin{equation}
E^{2}l^{2}(\varepsilon )-p^{2}h^{2}(\varepsilon )=m^{2},
\end{equation}%
where $\varepsilon =E/E_{P}$ is the ratio of energy $E$ of a particle with
mass $m$ to the Planck energy $E_{P}$. Also, $l(\varepsilon )$ and $%
h(\varepsilon )$ refer to rainbow functions that depend on energy. The
rainbow functions must equal unit whenever $\varepsilon $ tends to zero,
i.e., $\underset{\varepsilon \longrightarrow 0}{\lim }l(\varepsilon )=1$,
and $\underset{\varepsilon \longrightarrow 0}{\lim }h(\varepsilon )=1$. On
energy-dependent spacetime can be obtained in terms of orthonormal frame
fields 
\begin{equation}
H(\varepsilon )=\eta ^{ab}e_{a}(\varepsilon )\otimes e_{b}(\varepsilon ).
\end{equation}%
So, the energy-dependent linear element of a static spherical symmetric
spacetime in $(3+1)$ dimensions is given 
\begin{equation}
ds^{2}=-\frac{e^{2\nu \left( r\right) }}{l^{2}\left( \varepsilon \right) }%
dt^{2}+\frac{e^{2\lambda \left( r\right) }}{h^{2}\left( \varepsilon \right) }%
dr^{2}+\frac{r^{2}}{h^{2}\left( \varepsilon \right) }\left( d\theta
^{2}+\sin ^{2}\theta d\varphi ^{2}\right) ,  \label{metric}
\end{equation}%
where the metric potentials $e^{2\nu \left( r\right) }$ and $e^{2\lambda
\left( r\right) }$ have radial-dependent. In RG, the equations of motion are
given as follows 
\begin{equation}
G_{ab}(\varepsilon )=\dfrac{8\pi G(\varepsilon )}{c(\varepsilon )^{4}}%
T_{ab}\left( \varepsilon \right) ,  \label{GReq}
\end{equation}%
where $G_{ab}(\varepsilon )$ refers to energy-dependent Einstein's tensor.
Also, $G(\varepsilon )$ is an energy-dependent Newton's constant. This
energy-dependent of $G(\varepsilon )$ indicates that the effective
gravitational coupling depends on the energy and confirms a normalization
group equation \cite{Magueijo 2004}. Besides, $c(\varepsilon )$ is the
energy-dependent speed of light. Anyway, here we consider $G(\varepsilon
)=c(\varepsilon )=1$. The stress-energy tensor $T_{ab}\left( \varepsilon
\right) $ acts as an energy-dependent source. We consider that the interior
of the star is filled with an anisotropic fluid. The energy-momentum tensor
for an anisotropic distribution $T_{ab}$, is applied according to the
following definition \cite{Bayin1986} 
\begin{equation}
T_{ab}\left( \varepsilon \right) =\left[ \rho \left( r\right) +p_{t}\left(
r\right) \right] u_{a}u_{b}+p_{t}\left( r\right) g_{ab}+\left[ p_{r}\left(
r\right) -p_{t}\left( r\right) \right] x_{a}x_{b},  \label{T4}
\end{equation}%
where $\rho $ is the energy density. Also, $p_{r}$, and $p_{t}$ are the
radial pressure and the transverse pressure, respectively. The four-velocity
vector $u_{a}$ applies to normalization relation $u_{a}u^{a}=-1$, and it is
obtained from relation $u^{a}=l(\varepsilon )e^{-\nu (r)}\delta _{0}^{a}$.
The unit spacelike vector $x_{a}$ also is normalized by $x_{a}x^{a}=1$, and
it is determined by $x^{a}=h(\varepsilon )e^{-\lambda (r)}\delta _{1}^{a}$.
By substituting the metric in Eq. (\ref{T4}), the mixed diagonal elements of
the energy-momentum tensor are determined 
\begin{equation}
T_{a}^{b}=diag\left[ -\rho \left( r\right) ,p_{r}\left( r\right)
,p_{t}\left( r\right) ,p_{t}\left( r\right) \right] .  \label{T}
\end{equation}

To obtain the equations of motion, we put Eq. (\ref{T}) in Eqs. (\ref{GReq}%
). Therefore, the equations of motion are obtained in their extended form in
terms of rainbow functions, 
\begin{eqnarray}
&&\dfrac{h^{2}(\varepsilon )(2r\lambda ^{\prime }e^{-2\lambda }-e^{-2\lambda
}+1)}{r^{2}}=8\pi \rho ,  \label{GR1} \\
&&\dfrac{h^{2}(\varepsilon )(2r\nu ^{\prime }e^{-2\lambda }+e^{-2\lambda }-1)%
}{r^{2}}=8\pi p_{r},  \label{GR2} \\
&&h^{2}(\varepsilon )\left[ \nu ^{\prime \prime }-\lambda ^{\prime }\nu
^{\prime }+\nu ^{\prime 2}-\frac{\lambda ^{\prime }}{r}+\frac{\nu ^{\prime }%
}{r}\right] e^{-2\lambda }=8\pi p_{t},  \label{GR3}
\end{eqnarray}
where the prime and double prime display the first and second derivatives
with respect to $r$, respectively.

The "$tt$" component of the field equations, Eq. (\ref{GR1}), gives the
following relation 
\begin{equation}
e^{-2\lambda }=1-\dfrac{2M_{eff}\left( r,\varepsilon \right) }{r},
\label{lambda}
\end{equation}%
in which $M_{eff}\left( r,\varepsilon \right) =\int\limits_{0}^{r}\frac{4\pi
r^{\prime 2}\rho dr^{\prime }}{h^{2}\left( \varepsilon \right) }=\frac{%
m\left( r\right) }{h^{2}\left( \varepsilon \right) }$. It is notable that $%
M_{eff}\left(r,\varepsilon \right) $ represents the mass-energy function in
the RG, and it is known as effective mass, and also $m\left( r\right) $
refers to the mass-energy function in the GR. According to the "$rr$"
component of the field equations (\ref{GR2}), the gravity profile is
obtained, 
\begin{equation}
\frac{d\nu }{dr}=\frac{M_{eff}\left( r,\varepsilon \right) h^{2}\left(
\varepsilon \right) +4\pi r^{3}p_{r}}{r\left( r-2M_{eff}\left( r,\varepsilon
\right) \right) h^{2}\left( \varepsilon \right) }.  \label{nu}
\end{equation}

By inserting Eq. (\ref{nu}) into the conservation relation $\triangledown
^{a}T_{ab}=0$, and doing some calculations, the modified hydrostatic
equilibrium equation (which is known as modified TOV equation) for an
anisotropic fluid in the RG is determined \cite{Tudeshki2022}, 
\begin{equation}
\frac{dp_{r}}{dr}=-\frac{\left( 4\pi r^{3}p_{r}+h^{2}\left( \varepsilon
\right) M_{eff}\left( r,\varepsilon \right) \right) \left[ p_{r}+\rho \right]
}{r\left( r-2M_{eff}\left( r,\varepsilon \right) \right) h^{2}\left(
\varepsilon \right) }+\frac{2\left[ p_{t}-p_{r}\right] }{r}.  \label{ModTOVI}
\end{equation}%
It should be noted that for $h(\varepsilon )=1$, the modified TOV equation
(Eq. (\ref{ModTOVI})) reduces to the TOV equation in GR \cite%
{Tolman1939,Oppenheimer1939}.

There are various phenomenological motivations for choosing rainbow
functions $l(\varepsilon )$ and $h(\varepsilon )$. These functions are
mainly divided into three categories, depending on $\varepsilon $. One is
based on the loop quantum gravity theory \cite{Jacob2010,Amelino2013} in the
form of $l(\varepsilon )=1$ and $h(\varepsilon )=\sqrt{1-\eta \varepsilon
^{n}}$. The other is based on the hard spectra from gamma-ray bursts \cite%
{Amelino1998} as $l(\varepsilon )=\dfrac{e^{\beta \varepsilon }-1}{\beta
\varepsilon }$ and $h(\varepsilon )=1$. Another option that we are
interested to use in this study for rainbow functions, is to ensure that the
speed of light does not change \cite{Magueijo 2002}, which is defined $%
l(\varepsilon )=h(\varepsilon )=\frac{1}{1-\psi \varepsilon }$. Above, $\eta 
$, $\beta $, and $\psi $ are parameters determined by empirical methods. Due
to the positive assumption of rainbow functions, $\psi $ applies in $\psi <%
\dfrac{1}{\varepsilon }$. On the other hand, the maximum energy assumed for
a particle cannot exceed the Planck energy limit. So it must be $\varepsilon
<1 $. As a result, rainbow functions assume positive values greater than
unity, and they become $1$ in the IR limit (or for GR).

\section{Properties of DESs in RG and Observational Data}

\label{sec3} \textbf{EOS and Anisotropy Model:} In order to characterize the
internal behavior of the DES, an appropriate EOS must be considered. As a
suitable candidate for dark energy that is consistent with the observational
results such as CMB data, supernova data, etc., we can mention the
generalized Chaplygin gas (GCG) model, which is known as a cosmic fluid and
has an EOS; $p_{r}=-\hat{B}/\rho ^{\omega }$ \cite{Gorini2003,Xu2012}, where 
$\hat{B}$ is a constant with units of [length$^{-4}$], and also $\omega $ is
a constant in the range of $0\leq \omega \leq 1$. By adding a barotropic
term to GCG EOS, the modified Chaplygin gas (MCG) EOS is defined as follows 
\cite{Debnath2004,Benaoum2012,Mazumder2012}, 
\begin{equation}
p_{r}=\hat{A}\rho -\frac{\hat{B}}{\rho ^{\omega }},  \label{EOS}
\end{equation}%
where the constant dimensionless values of $\hat{A}$ are determined from the
considerations of establishing the causality condition at the star surface.
The above equation can be written in the following form \cite%
{Panotopoulos2020,Panotopoulos2021}, 
\begin{equation}
p_{r}=A^{2}\rho -\dfrac{B^{2}}{\rho }.  \label{EOS1}
\end{equation}%
Here $\omega =1$ is assumed. The square of sound speed should be less than $1
$, i.e., $0<V_{sr}^{2}=\dfrac{dp_{r}}{d\rho} =A^{2}+B^{2}/ \rho^{2}<1$,
because the sound speed should be less than the speed of light. So the
causality condition must be satisfied. On the other hand, at the surface of
star $(r=R)$, the radial pressure is neglected, so the energy density and
constants $A$ and $B$ apply in the relation $B=A\rho$. As a result, the
causality condition is obeyed by $V_{sr}^{2}(R)=2A^{2}<1$. It is clear that $%
{A}$ must be less than $\sqrt{0.5}$ and $B$ is determined using values of $A 
$ \cite{Pretel2023,Tudeshki2023}. According to above discussions, in this
paper, we use the extended Chaplygin EOS (\ref{EOS1}) to make a comparison
with the previous studies \cite%
{Panotopoulos2020,Panotopoulos2021,Tudeshki2023}, where values $A=\sqrt{0.4}$
and $B=0.2\times10^{-3}km^{-2}$ are included in the numerical solution.

One of the unknown quantities in the hydrostatic equilibrium equation is the
anisotropy factor $\Delta $, which includes radial and transverse pressures
in the form of $p_{t}-p_{r}$. In examining more realistic models of stars,
it has been shown that in densities greater than nuclear density, in
addition to radial pressure inside the star, transverse pressure is also
created perpendicular to radial direction \cite{Ruderman,Canuto}. In
addition, the occurrence of physical phenomena such as phase transition \cite%
{Sokolov}, condensation \cite{Hartle} and electromagnetic fields \cite%
{Usov,bordbar2022} are also effective in creating anisotropy. In some
studies \cite%
{Bowers1974,Cosenza1981,Horvat2010,Doneva2012,Herrera2013,Raposo2019},
several models have been presented to describe the behavior of the
anisotropy factor. Here, inspired by the nonlinear model presented by Bowers
and Liang \cite{Bowers1974} (called BL anisotropic model), we define a
modified version of this model in the RG, which is as follows, 
\begin{equation}
\Delta =p_{t}-p_{r}=\dfrac{\Upsilon \left( \rho +3p_{r}\right) \left( \rho
+p_{r}\right) r^{2}}{1-\frac{2M_{eff}\left( r,\varepsilon \right) }{r}},
\label{delta}
\end{equation}%
where $\Upsilon $ is a constant determining the degree of anisotropy. $%
\Upsilon $ must be less than $2/3$ to satisfy condition $%
(M_{eff}/R)_{critical}>0$ \cite{Bowers1974}.

\textbf{Numerical Solutions:} The numerical solution of three coupled
differential equations (\ref{nu}), (\ref{ModTOVI}), and effective mass for
two cases $\Upsilon =0$ and $\Upsilon \neq 0$ leads to the determination of
the properties of DESs in RG with two isotropic and anisotropic
configurations, respectively. Numerical solution is made by considering the
following initial conditions at the center of the star, $M_{eff}(r=0)=0$,
and $\rho(r=0)=\rho _{c}$. Also, the boundary condition on the surface $%
p_{r}(R)=0$, in addition to $M(R)=M_{eff}(r=R)$, and $\nu (R)=\dfrac{1}{2}%
ln\left(1-2M(R)/R\right)$, where $\rho _{c}$ is the central energy density, $%
M(R)$ is the total mass of DES in RG and $\nu (R)$ is related to metric
potentials at radius $R$. It should be noted that conditions $\Delta (r=0)=0$
and $\rho _{s}=B/A$ are also met. In the following, the results for the
central density $\rho _{c}=1.35\times10^{15}~g/cm^{3}$ and different values
of the rainbow function $h(\varepsilon )$ are specified in Tables \ref{tab1}
and \ref{tab2} for isotropic and anisotropic fluid. Also, for a range of
central energy density, the maximum mass and its corresponding radius are
obtained.

\begin{table}[h]
\caption{The properties of isotropic DES in RG with $A=\protect\sqrt{0.4}$, $%
B=0.2\times 10^{-3}$ and different values of $h(\protect\varepsilon )$.}
\label{tab1}
\begin{center}
\begin{tabular}{|c|c|c|c|c|c|}
\hline
$h(\varepsilon)$ & $M_{max}[M_{_{\odot}}]$ & $R[km]$ & $R_{Sch}[km]$ & ~~ $%
\sigma$ ~~ & ~~ $Z_{s}$ ~~ \\ \hline\hline
$1$ & $2.64$ & $12.63$ & $7.83$ & $0.30$ & $0.61$ \\ \hline
$1.15$ & $3.04$ & $14.53$ & $9.00$ & $0.30$ & $0.61$ \\ \hline
$1.20$ & $3.17$ & $15.16$ & $9.39$ & $0.30$ & $0.61$ \\ \hline
$1.25$ & $3.31$ & $15.79$ & $9.78$ & $0.30$ & $0.61$ \\ \hline
$1.30$ & $3.44$ & $16.42$ & $10.18$ & $0.30$ & $0.61$ \\ \hline
$1.35$ & $3.57$ & $17.06$ & $10.57$ & $0.30$ & $0.61$ \\ \hline
$1.40$ & $3.70$ & $17.69$ & $10.96$ & $0.30$ & $0.61$ \\ \hline
\end{tabular}%
\end{center}
\end{table}

\begin{table}[h]
\caption{The properties of anisotropic DES in RG with $A=\protect\sqrt{0.4}$%
, $B=0.23\times 10^{-3}$, $\Upsilon =0.1$ and different values of $h(\protect%
\varepsilon )$.}
\label{tab2}
\begin{center}
\begin{tabular}{|c|c|c|c|c|c|c|}
\hline
$h(\varepsilon)$ & $M_{max}[M_{_{\odot}}]$ & $R[km]$ & $R_{Sch}[km]$ & $%
\Delta_{s}[dyn/cm^{2}]$ & ~~ $\sigma$ ~~ & ~$Z_{s}$ ~ \\ \hline\hline
$1$ & $2.72$ & $12.83$ & $8.07$ & $5.37\times10^{33}$ & $0.31$ & $0.64$ \\ 
\hline
$1.15$ & $3.16$ & $14.80$ & $9.38$ & $7.24\times10^{33}$ & $0.31$ & $0.65$
\\ \hline
$1.2$ & $3.19$ & $15.46$ & $9.82$ & $7.93\times10^{33}$ & $0.31$ & $0.65$ \\ 
\hline
$1.25$ & $3.47$ & $16.12$ & $10.27$ & $8.67\times10^{33}$ & $0.31$ & $0.65$
\\ \hline
$1.3$ & $3.62$ & $16.78$ & $10.72$ & $9.44\times10^{33}$ & $0.31$ & $0.66$
\\ \hline
$1.35$ & $3.77$ & $17.45$ & $11.18$ & $1.02\times10^{34}$ & $0.32$ & $0.66$
\\ \hline
$1.4$ & $3.93$ & $18.11$ & $11.64$ & $1.11\times10^{34}$ & $0.32$ & $0.67$
\\ \hline
\end{tabular}%
\end{center}
\end{table}

\begin{figure}[tbh]
\centering
\includegraphics[width=0.4\textwidth]{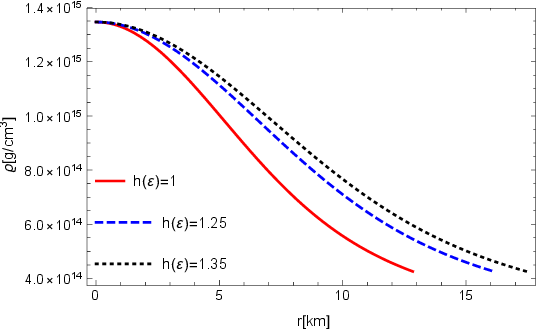} %
\includegraphics[width=0.4\textwidth]{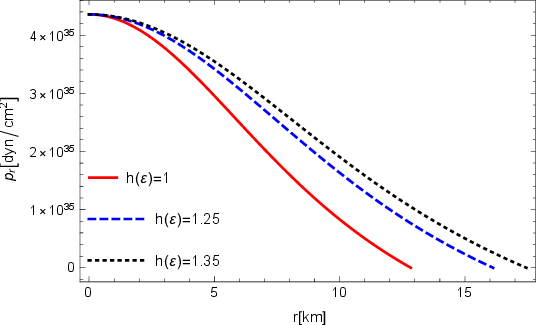}
\caption{ Density $\protect\varrho $ vs radius $r$ (left panel) and radial
pressure $p_{r}$ vs radius $r$ for different $h(\protect\varepsilon )$
(right panel) with $\Upsilon =0.1$, $A=\protect\sqrt{0.4}$ and $B=0.2\times
10^{-3}$. }
\label{Fig1}
\end{figure}

\textbf{Pressure, Density and Anisotropy Factor:} The behaviors of density
and radial pressure versus distance are plotted in Fig. \ref{Fig1}. As one
can see, the density (pressure) in the center of DES has the maximum value.
As it moves toward the surface of DES, it finds a downward trend, and
finally reaches its minimum value at the surface. Increasing the rainbow
function $h(\varepsilon )$ at a certain density results in increasing the
radius. 

\begin{figure}[tbh]
\centering
\includegraphics[width=0.4\textwidth]{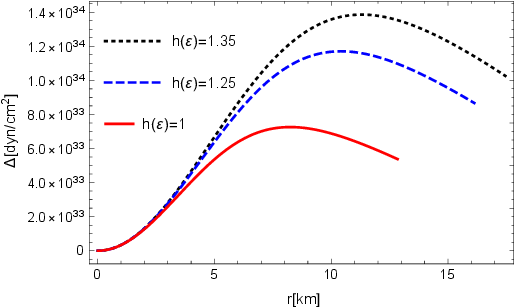} %
\includegraphics[width=0.4\textwidth]{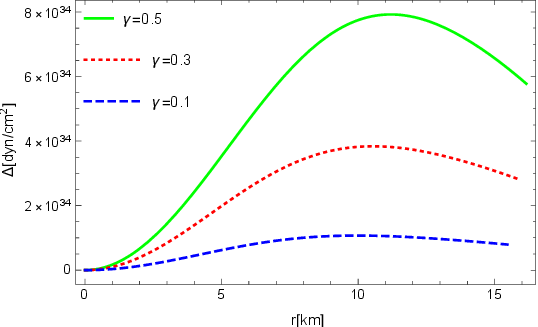}
\caption{Anisotropy factor ($\Delta $) vs radius $r$ for $\Upsilon =0.1$ and
different $h(\protect\varepsilon )$ (left panel). For $h(\protect\varepsilon %
)=1.20$, $A=\protect\sqrt{0.4}$, $B=0.2\times 10^{-3}$ and different $%
\Upsilon $ (right panel). }
\label{Fig2}
\end{figure}

Fig. \ref{Fig2} shows the anisotropy factor $\Delta $ versus radius. In the
left panel of Fig. \ref{Fig2}, we see that by increasing the value of
rainbow function $h(\varepsilon)$, the $\Delta $ value has also increased.
In the right panel of Fig. \ref{Fig2}, it can be seen that the $\Delta $
increases by increasing the degree of anisotropy $\Upsilon $. Note that in
the center of the star $\Delta =0$, because $p_{t}(r=0)=p_{r}(r=0)$. 

\textbf{Maximum Mass and its Corresponding Radius:} There are Several
reasons to make the maximum mass calculated for compact objects such as
neutron stars approaches the result obtained from the GW190814 event $%
(M_{max}>2.5M_{\odot })$, including the presence of dense matter with
different EOSs. It has also shown that the presence of anisotropy is also
able to increase the maximum mass in a DES \cite{Pretel2023}. However, due
to the unknown nature of dark energy, various EOSs with free parameters are
consistent with observational results, capable of predicting the mass range
region of DES \cite{Panotopoulos2020,Panotopoulos2021}. The results have
shown that in addition to the fact that the maximum mass calculated for the
DES is within the mass range obtained for the neutron stars, it is even
capable of justifying the mass gap region between the massive neutron star
and the low-mass black hole $2M_{\odot }-5M_{\odot }$ \cite%
{Abbott2020,Ozel,Thompson2019}. Despite the study on the structure of
compact stars with dark energy in GR \cite{Haghani}, however, DESs have been
studied in modified gravity such as massive gravity and suggested a maximum
mass of more than $3M_{\odot }$ \cite{Tudeshki2023}. Motivated by the effect
of high-energy limit (UV limit) on compact objects such as DES, in present
work, we investigate the effect of the energy-dependent rainbow function $%
h(\varepsilon )$ on the maximum mass $M_{max}$ and its corresponding radius $%
R$ for two cases of isotropic and anisotropic fluid using a numerical
solution. Since we intend to study the role of RG and the anisotropy
parameter $\Upsilon $ on the behavior of the maximum mass $M_{max}$ of DES,
we fix other involved parameters, including the parameters of EOS $A$ and $B$%
.

\begin{figure}[tbh]
\centering
\includegraphics[width=0.5\textwidth]{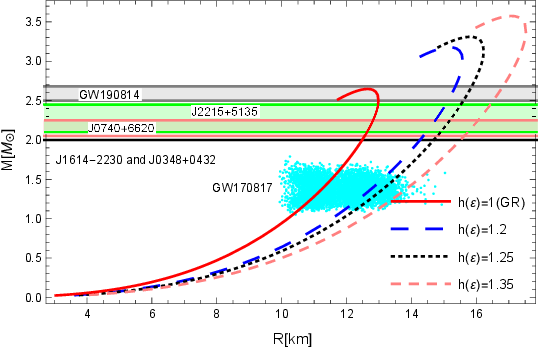} \newline
\includegraphics[width=0.5\textwidth]{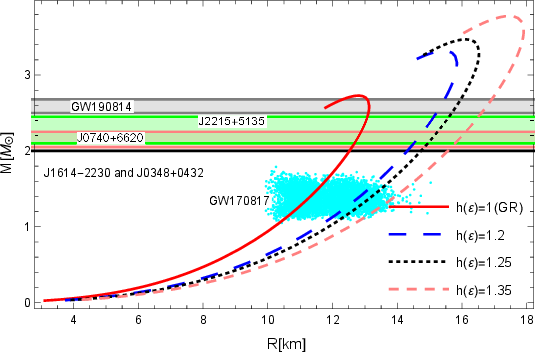} \newline
\includegraphics[width=0.5\textwidth]{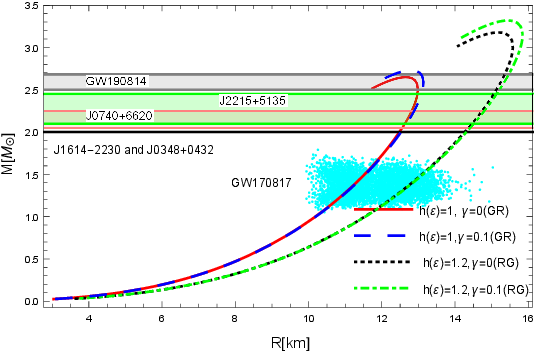} \newline
\caption{Mass-radius relation. Isotropic model $\Upsilon =0$ (up panel),
anisotropic model $\Upsilon \neq 0$ (middle panel) for DES with $A=\protect%
\sqrt{0.4}$ and $B=0.2\times 10^{-3}$ and different values $h(\protect%
\varepsilon )$. Comparison of isotropic and anisotropic models in GR and RG
(down panel). At $M=2M_{\odot }$, the black horizontal line shows the mass
data of the massive NS pulsars (J1614-2230 and J0348+0432). The wide pink
and green bars display the observational mass data of NS pulsars J0740+6620
and J2215+5135, respectively. Observational data from event GW190814 has
been shown in the gray bar. Also the cyan area shows mass-radius constraints
of component binary pair mass-radius constraints (GW170817 event).}
\label{Fig3}
\end{figure}

Mass-radius relation diagrams in Fig. \ref{Fig3} contain colored bands,
where each of them represents the mass range of observational candidates,
which are introduced in the caption below the diagram. According to the
upper panel in Fig. \ref{Fig3}, the curve related to isotropic GR passes
through all color bands, and its maximum reaches that of GW190814. But other
curves related to isotropic RG, with the increase of the rainbow function $%
h(\varepsilon )$ up to $1.35$ also exceed the limit of GW190814, and the
maximum mass reaches the value of $3.7M_{\odot }$. In the middle panel in
Fig. \ref{Fig3}, the anisotropic GR curve is slightly above the limit of
GW190814. Also, in other curves, the maxima in RG increase with the presence
of $\Upsilon$ and reach the highest value in $3.9M_{\odot }$. With a look at
the bottom panel of Fig. \ref{Fig3}, one can see that although GR covers the
observational data range NS pulsars (J1614-2230 and J0348+0432), NS pulsars
J0740+6620 and J2215+5135 and also GW190814, but compared to RG, it is not
able to cover the areas above the gray band. These regions include mass
range 2MASS J05215658+4359220 and GW190425 with values $3.3M_{\odot }$ and $%
3.4M_{\odot }$, respectively, which are satisfied only in RG (For
simplicity, observational data J05215658+4359220 and GW190425 are not
included in the plot). An interesting point is that even if we increase the
degree of anisotropy of the model in GR, the anisotropic RG model provides
us with larger values of the maximum mass. At the end of this part, we
display the agreement of the results obtained for the RG and GR frameworks
with the data of different observational candidates in Table. \ref{tab3}. It
should be noted that this comparison is made for the fixed values $A=\sqrt{%
0.4}$, $B=0.2\times 10^{-3}$, and $\Upsilon =0.1$. 
\begin{table}[tbh]
\caption{Different observational data and agreement with the calculation
results of Chaplygin EOS between general relativity (GR) and gravity's
rainbow (RG). It is notable that the abbreviations of "iso" and "ani" are
related to the isotropic and anisotropic, respectively.}
\label{tab3}
\begin{center}
\begin{tabular}{|c|c|c||c|c|}
\hline
Observational & GR & GR & RG & RG \\ 
candidate & (iso) & (ani) & (iso) & (ani) \\ \hline\hline
GW190814 & $\checkmark$ & $\checkmark$ & $\checkmark$ & $\checkmark$ \\ 
\hline
GW170817 & $\checkmark$ & $\checkmark$ & $\checkmark$ & $\checkmark$ \\ 
\hline
GW190425 & $\times$ & $\times$ & $\checkmark$ & $\checkmark$ \\ \hline
J05215658+4359220 & $\times$ & $\times$ & $\checkmark$ & $\checkmark$ \\ 
\hline
PSR J1614-2230 & $\checkmark$ & $\checkmark$ & $\checkmark$ & $\checkmark$
\\ \hline
PSR J0348+0432 & $\checkmark$ & $\checkmark$ & $\checkmark$ & $\checkmark$
\\ \hline
PSR J0740+6620 & $\checkmark$ & $\checkmark$ & $\checkmark$ & $\checkmark$
\\ \hline
PSR J2215+5135 & $\checkmark$ & $\checkmark$ & $\checkmark$ & $\checkmark$
\\ \hline
PSR J1311-3430 & $\checkmark$ & $\checkmark$ & $\checkmark$ & $\checkmark$
\\ \hline
PSR J1748-2021B & $\checkmark$ & $\checkmark$ & $\checkmark$ & $\checkmark$
\\ \hline
\end{tabular}%
\end{center}
\end{table}

\textbf{Schwarzschild Radius}: To calculate the modified Schwarzschild
radius $R_{Sch}$ in RG, we must set the metric function (\ref{lambda}) equal
to zero, i.e., $e^{-2\lambda (r=R_{Sch})}=0$. So, the modified Schwarzschild
radius in the RG is obtained \cite%
{HendiFaizal2016,HendiPanahiyan2016,HendiBordbar2016}, 
\begin{equation}
R_{Sch} =2M_{eff}.  \label{Rsch RG}
\end{equation}
Since the effective mass $M_{eff}$ depends on the rainbow function $%
h(\varepsilon)$, as a result, $R_{Sch}$ also depends on the energy.
According to the results of Tables \ref{tab1} and \ref{tab2}, we see that by
increasing the rainbow function $h(\varepsilon)$, $R_{Sch}$ also increases.
Notably, these compact objects cannot be black holes, because their radii
are less than Schwarzschild radii ($R>R_{Sch}$).

\textbf{Compactness:} The degree of compactness of a compact object is
defined by the ratio of its mass $M$ to its radius $R$. The mass obtained in
the gravity's rainbow is defined in terms of effective mass $M_{eff}$.
Therefore, the compactness relation in RG is generalized as follows, 
\begin{equation}
\sigma =\dfrac{M_{eff}}{R}.  \label{comp}
\end{equation}%
According to the results of Tables \ref{tab1} and (\ref{tab2}), it can be
seen that for different values of the rainbow function $h(\varepsilon )$,
the compactness remains almost unchanged in both isotropic and anisotropic
models.

\begin{figure}[tbh]
\centering
\includegraphics[width=0.5\textwidth]{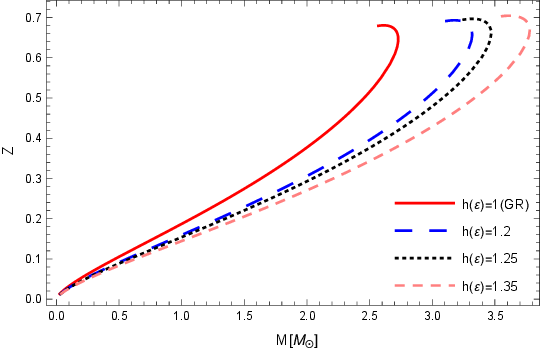} \newline
\caption{ Surface gravitational redshift ($Z_{s}$) vs mass ($M$) in the RG
with different $h(\protect\varepsilon )$ for anisotropic model with $A=%
\protect\sqrt{0.4}$, and $B=0.2\times 10^{-3}$.}
\label{Fig5}
\end{figure}

\textbf{Gravitational Redshift:} We can obtain the surface gravitational
redshift $Z_{s}$ using Eq. (\ref{lambda}) and definition $1+Z_{s}=e^{\lambda
(R)}$ as follows \cite{HendiBordbar2016}, 
\begin{equation}
1+Z_{s}=\dfrac{1}{\sqrt{1-2\sigma }}.  \label{redshift}
\end{equation}
As can be seen from the above relation, surface gravitational redshift is
related to compactness. Since the $\sigma $ is approximately fixed for
different values of the rainbow function $h(\varepsilon )$, as a result, $%
Z_{s}$ also has very small changes (see Tables \ref{tab1}, and \ref{tab2},
and also Fig. \ref{Fig5}). 

\textbf{Energy Condition}: For an anisotropic fluid, the energy conditions
are valid if all conditions such as the null energy condition (NEC), weak
energy condition (WEC), strong energy condition(SEC), and dominant energy
condition (DEC) are satisfied \cite{Leon1993}. Using the numerical results
obtained for energy density $\rho $, radial pressure $p_{r}$ and anisotropy
factor $\Delta =p_{t}-p_{r}$, according to the Table \ref{tab4}, these
conditions are briefly categorized, and their validity is evaluated. It can
be seen that all the energy conditions are valid in the RG. Although MCG EOS
describes the dark energy in this star, the strong energy condition is
satisfied, which indicates that the dark energy in this star behaves like
matter.

\begin{table}[h]
\caption{Energy conditions of DES in RG with $h(\protect\varepsilon)=1.2$, $%
A=\protect\sqrt{0.4}$, $\Upsilon=0.1$, and $B=0.2\times10^{-3}$.}
\label{tab4}
\begin{center}
\begin{tabular}{|c|c|c|c|}
\hline
NEC & WEC & SEC & DEC \\ \hline\hline
$\rho +p_{r}\geq 0$ & $\rho \geq 0$ & $\rho+p_{r}+2p_{t}\geq 0$ & $\rho \geq
\left\vert p_{r}\right\vert $ \\ 
$\rho +p_{t}\geq 0$ & $\rho +p_{r}\geq 0$,~ $\rho +p_{t}\geq 0$ & $\rho
+p_{t}\geq 0$ & $\rho \geq \left\vert p_{t}\right\vert $ \\ \hline
$\checkmark$ & $\checkmark$ & $\checkmark$ & $\checkmark$ \\ \hline
\end{tabular}%
\end{center}
\end{table}

\section{STABILITY}

\label{sec4} To check the stability of compact objects, it is customary to
use several different tests, which are mentioned below.

\subsection{Causality and Cracking}

\begin{figure}[tbh]
\centering
\includegraphics[width=0.4\textwidth]{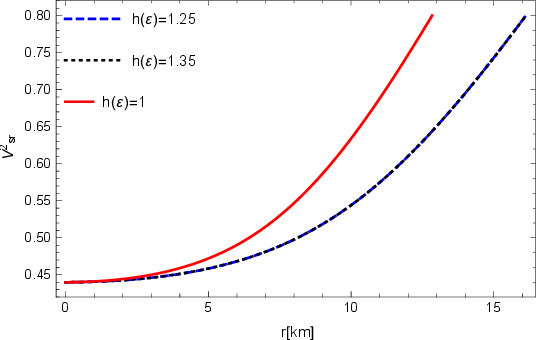} %
\includegraphics[width=0.4\textwidth]{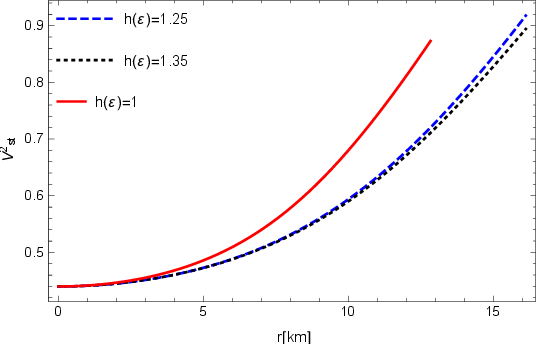} \newline
\caption{Square of radial speed of sound $V_{sr}^{2}$ (left panel) and
square of transverse speed of sound $V_{st}^{2}$ (right panel) vs radius in
RG for $A=\protect\sqrt{0.4}$, $B=0.2\times 10^{-3}$, $\Upsilon =0.1$, and
different $h(\protect\varepsilon)$.}
\label{Fig6}
\end{figure}

\begin{figure}[h]
\centering
\includegraphics[width=0.5\textwidth]{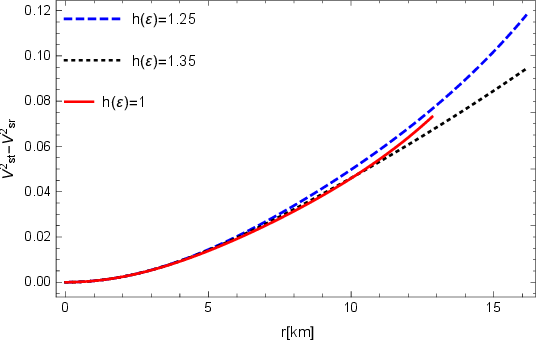}\newline
\caption{ $V_{st}^{2}-V_{sr}^{2}$ vs radius in RG for $A=\protect\sqrt{0.4}$%
, $B=0.2\times 10^{-3}$, $\Upsilon=0.1$, and different $h(\protect%
\varepsilon )$.}
\label{Fig7}
\end{figure}

In models of anisotropic fluid where the fluid has two radial and transverse
pressures, the causality condition is valid when two conditions, $%
0<V_{sr}^{2}=\dfrac{dp_{r}}{d\rho }<1$ and $0<V_{st}^{2}=\dfrac{dp_{t}}{%
d\rho }<1$ are met \cite{Herrera1992}. The quantities $V_{sr}$ and $V_{st}$
are the radial and transverse speeds of sound, respectively. This means that
the speed of sound in both radial and transverse directions should not
exceed the speed of light. In Fig. (\ref{Fig6}), $V_{sr}^{2}$ and $%
V_{st}^{2} $ are plotted versus radius. 
According to Fig. \ref{Fig6} for different values of $h(\varepsilon )$, the
causality condition is still satisfied. Also, an interesting phenomenon
occurs in locally anisotropic configurations, known as cracking \cite%
{Herrera1992,Abreu2007}. In fact, the ratio of the anisotropy factor
perturbations to density perturbations in a fluid creates cracking, which is
related to the sound speed in the following form, 
\begin{equation}
\dfrac{\delta \Delta }{\delta \rho }\sim \dfrac{\delta (p_{t}-p_{r})}{\delta
\rho }\sim \dfrac{\delta p_{t}}{\delta \rho }-\dfrac{\delta p_{r}}{\delta
\rho }\sim V_{st}^{2}-V_{sr}^{2}.  \label{craking}
\end{equation}%
This difference in the square of sound speed is a measure that shows the
state of stability of the compact object against perturbations caused by
anisotropy. Therefore, the following two criteria are used to check the
stability under cracking, 
\begin{equation}
-1<V_{st}^{2}-V_{sr}^{2}<0~~~\text{potentially~stable,}
\end{equation}%
\begin{equation}
0<V_{st}^{2}-V_{sr}^{2}<1~~~\text{potentially~unstable.}
\end{equation}%
Fig. (\ref{Fig7}) shows a view of the cracking criteria, i.e., the
difference between $V_{st}^{2}-V_{sr}^{2}$ versus the radius. As can be
seen, for different values of the rainbow function $h(\varepsilon )$, the
curves are in the positive part of the plot, and this means that the system
is potentially unstable. 

\subsection{Adiabatic Index}

\begin{figure}[tbh]
\centering
\includegraphics[width=0.5\textwidth]{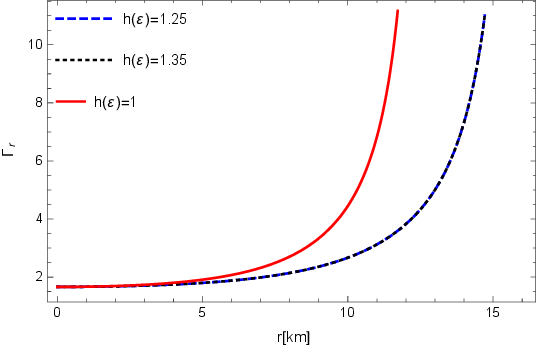}
\caption{Adiabatic index ($\Gamma _{r}$) vs radius in RG with $A=\protect%
\sqrt{0.4}$, $B=0.2\times 10^{-3}$, and different $h(\protect\varepsilon )$.}
\label{Fig8}
\end{figure}

Under the radial perturbations created in an anisotropic fluid, the
adiabatic index can be considered as follows \cite%
{Chandrasekhar1964,Bondi1964,Chan1993}, 
\begin{equation}
\Gamma _{r}=\left( 1+\dfrac{\rho }{p_{r}}\right) V_{sr}^{2}.
\end{equation}%
Although the system is under anisotropy, checking its dynamic stability
through the adiabatic index is sufficient for the above relation, because
these are radial perturbations. Regions are dynamically stable where the
adiabatic index is $\Gamma _{r}>1.33$. In Fig. \ref{Fig8}, the adiabatic
index curves are drawn for different values of the rainbow function. It can
be seen that all values are greater than $1.33$. Therefore, the obtained
DESs in the RG are dynamically stable. 

\subsection{Harrison-Zeldovich-Novikov condition}

\begin{figure}[!tbh]
\centering
\includegraphics[width=0.5\textwidth]{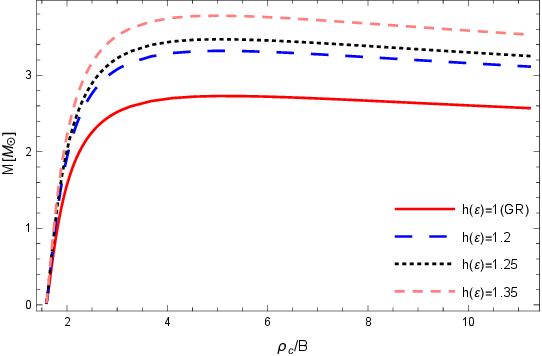}
\caption{Mass-central energy density relation with anisotropic model $%
\Upsilon\neq 0$ with $A=\protect\sqrt{0.3}$, $B=0.2\times 10^{-3}$, and
different values $h(\protect\varepsilon)$.}
\label{Fig4}
\end{figure}

Harrison-Zeldovich-Novikov condition \cite{Zeldovich1971,Harrison1965}
states that a compact object maintains its dynamical stability if it
satisfies the condition $dM/d\rho _{c}>0$. This means that as long as this
condition is met, the compact object resists gravitational collapse. In Fig. %
\ref{Fig4}, we see that the curves have an ascending trend up to a critical
energy density $\rho_{crit}$. Here, our studied system is considered up to $%
dM/d\rho_{c}=0$, where the critical density is $\rho_{crit} $. As a result,
the dark energy star is completely stable in the range $\rho_{c}\leq
\rho_{crit}$. After this limit, i.e. $\rho_{c}> \rho_{crit}$, the star
becomes unstable, and it may collapse and turn into a black hole.

\section{Summary and Conclusions}

\label{sec5} In summary, in this study, we discussed the possible effects of
energy-dependent spacetime on the structure of dark energy star and
especially the maximum mass $M_{max}$ and its corresponding radius $R$. For
this purpose, we obtained the generalized hydrostatic equilibrium equation
in the gravity's rainbow for an anisotropic fluid by using the extended
Chaplygin gas equation of state and Bowers and Liang anisotropy model. Then,
using the numerical solutions in two gravitational frameworks, general
relativity and gravity's rainbow, the behavior of physical quantities of the
dark energy star such as density, radial pressure and anisotropy, the
maximum mass and its corresponding radius, compactness, and gravitational
redshift for the different values of the rainbow function $h(\varepsilon )$
were analyzed in two isotropic and anisotropic models.

The obtained results showed that we can make a constraint on the maximum
mass of the dark energy star by gravity's rainbow. This dependence on the
energy leads the maximum mass $M_{max}\sim3.70$ in the isotropic model and
the maximum mass $M_{max}\sim3.90$ in the anisotropic model. These obtained
masses are in the mass gap region $2M_{\odot }-5M_{\odot }$, which could be
a justification for the existence of observed compact objects similar to the
mass observed in two observational candidates GW190425 and
J05215658+4359220. We also indicated that, in comparison between gravity's
rainbow and general relativity, gravity's rainbow could cover a wider range
of mass gap regions.

The stability of dark energy star was studied with different tests. It was
shown that although the causality condition in the configuration is
satisfied, the perturbations due to anisotropy induce a crack in the system,
which makes the dark energy star potentially unstable. However, the dark
energy star is stable up to a critical limit for the central density. Also
due to the increase in central density, the star will be undergo
gravitational collapse.

From a general point of view, it can be said that the dependence of
spacetime on the energy of particles has a significant effect on the maximum
mass of compact objects. Matching these results obtained from our
theoretical model with observational evidence will help us to improve our
other theoretical models in the future.

\begin{acknowledgements}
ABT and GHB wish to thank Shiraz University research council. BEP thanks the University of Mazandaran. The University of Mazandaran has supported the work of BEP by title "Evolution of the masses of celestial compact objects in various gravity".
\end{acknowledgements}

\end{document}